\documentclass[12pt,preprint]{aastex}
\usepackage{natbib,graphicx}

\shorttitle{Black Hole Spins}
\shortauthors{Daly}

\begin{document}
 
\title{Black Hole Spins of Radio Sources}
\author{Ruth A. Daly\altaffilmark{~}}
\affil{Department of Physics, Penn State University, Berks Campus, P. O. 
Box 7009, Reading, PA 19610}
\email{rdaly@psu.edu}

\begin{abstract}
A new approach to constraining or determining  
the spin of a massive black hole is proposed. A  
key parameter in the analysis is the dimensionless
ratio, $r$, of the energy released to the mass of the hole. 
It is shown that the black hole spin $j$ may be written as a function
of the dimensionless ratio of the black hole 
spin energy to the black hole mass.
When extraction of the black hole spin energy powers an 
outflow from the hole, 
the ratio $r$ provides an indication of the black hole spin. 
The method is applied to a 
sample of 19 very powerful classical double radio galaxies with 
a range of size and redshift. 
The sources are found to have remarkably similar 
values of $r$ and $j$, 
implying that the sources have very similar physical conditions
at the time the outflow is generated.  The 
weighted mean value of the spins is $0.12 \pm 0.01$.
A sample of 29 central dominant
galaxies is also studied; 
most of these sources have amorphous or FRI radio structure. 
The sources have a broad range of 
ratios $r$, and correspondingly broad range of black hole 
spins $j$ with values 
from about 0.001 to 0.4, and a median value of about 0.03. 
The broad range probably results from the fact that each source is 
observed at a different stage in its lifetime.  The intrinsic range
of parameters could be as tight as it is for the powerful classical 
double radio galaxies.

\end{abstract}

\keywords{black hole physics --- galaxies: active --- galaxies: nuclei}

\section{INTRODUCTION}

Supermassive black holes are believed to 
power quasars and other types of AGN.
The AGN activity manifests itself as radiation from the 
region around the black hole and as highly collimated outflows from 
the immediate vicinity of the black hole (e.g. Rees 1984).  
Two defining properties of astrophysical black 
holes that can be measured in principle are the black hole mass and spin.
A significant
amount of progress has been made toward measuring the masses of
supermassive black holes (e.g. Kormendy \& Richstone 1995; 
Ferrarese \& Ford 2005), though measuring the spin 
has been more challenging. 

General theoretical studies of black hole spins 
indicate that accretion of gas 
is likely to produce rapidly spinning 
black holes (Volonteri et al. 2005), the merger of two nonspinning
black holes of comparable mass is likely to produce a rapidly spinning
black hole (Gammie, Shapiro, \& McKinney 2004), 
and the capture of a smaller
companion may 
cause the hole to spin down (Hughes \& Blandford 2003). 
To date, there are only a few direct
observations that allow black hole spins to be studied.
Observations of Seyfert galaxies suggest rapidly rotating  
black holes in these systems (Wilms et al. 2001; 
Fabian et al. 2002).  A large spin is indicated by 
observations of the Galactic Center black
hole (Genzel et al. 2003; Ashenbach et al. 2004). 
And, X-ray observations of active galaxies suggest
rapidly rotating black holes in these systems (Crummy et al. 2006). 

Theoretical predictions for the spins of black holes for FRI and
FRII radio sources (defined by Fanaroff \& Riley 1974) have been
made by Meier (1999) in the context of a 
rotating black hole model, and several predictions of this
model are very similar to those obtained in the model of 
Blandford \& Znajek (1977). 

In this work,  
a new method of measuring the spin of a supermassive black hole is
proposed. The method may be applied to sources with large-scale
outflows that power radio sources, and, perhaps, other
types of sources. 
A double radio 
source is powered by two oppositely directed highly collimated 
outflows from the vicinity of a supermassive black hole. 
The spin energy that is available to power the outflow can be
written in terms of the black hole mass and spin. 
Thus, if the spin energy and black hole mass can be
estimated or constrained, the black hole spin can be estimated
or constrained. 

The method to obtain a measure of the black hole spin 
is described in section 2. This method is independent of the 
specific model of energy extraction.  
The results obtained
by applying the method are presented in section 3. 
A discussion follows in section 4. 
The results are obtained assuming a standard 
cosmological
model with $H_o = 70$ km/s/Mpc, $\Omega_m = 0.3$, $\Omega_{\Lambda}=0.7$,
and zero space curvature.

\section{THE METHOD}

The energy $E_*$ that can be extracted from a black hole
is related to the black hole spin $j \equiv (a/m)$ and mass $M$, 
as discussed, for example, by Blandford (1990), where
$a$ is defined in terms of the spin angular momentum $S$, the 
speed of light $c$ and the black hole mass $M$, $a \equiv S/(Mc)$,
and $m$ is the gravitational radius of the black hole, $m = GM/c^2$. 
The spin energy is $(M-M_i)c^2$, where $M_i$ is the irreducible mass of the 
hole.  Setting $E_*$ equal to 
the spin energy and solving for the 
black hole spin, we find that 
\begin{equation}
j =  2~ (2r -5r^2+4r^3-r^4)^{1/2}
\end{equation}
where $r \equiv E_*/(c^2M)$, or
$j \approx (8r)^{1/2}$ for $r \ll 1$.  
Thus, for sources with outflows powered by the spin energy of a black hole,
the black hole spin $j$ can be measured when the total energy $E_*$ of the
outflow and the black hole mass $M$ are known. 
Determinations of the black hole spin obtained using equation (1) do not
depend upon the details of the energy extraction.  The use of 
this equation assumes only that
the ultimate source of the energy is the black hole spin, and that, 
during the short period of spin energy extraction, the black hole
spin changes only because of the energy extraction.    

The ratio $r$ provides an important diagnostic of the black hole system.  
Studies of the ratio $r$ provide insight into the physical state
of the system at the time the outflow is generated irrespective of 
whether the outflow is powered by the spin energy of
the hole. 

\section{RESULTS}
\subsection{Results Obtained with Powerful Classical Double Radio Galaxies}
\label{FRII}

The spin energy of the black hole is thought to power the 
large-scale outflows from AGN (e.g. Blandford \& Znajek 1977; 
Rees 1984; Blandford 1990; Wilson \& Colbert 1995; Daly 1995; 
Meier 1999, 2002).
The total energy $E_*$ that will be channeled through a large-scale 
outflow can be determined for the most powerful classical double 
radio galaxies (e.g. Daly \& Guerra 2002;
Wan, Daly, \& Guerra 2002; Daly et al. 2008; O'Dea et al. 2008). 
This empirically determined total energy $E_*$ can 
then be combined with the black hole mass to obtain the ratio $r$ and 
a measure of 
or bound on the spin parameter of sources using eq. (1).  
If only part of the spin energy is tapped, or if some of the tapped
energy is dissipated or radiated away 
and does not reach the extremities of the radio source,
the value of $j$ obtained using eq. (1) will be a lower bound. 
If the outflows are powered by energy from some other source, 
such as an accretion disk, then the energy per unit black hole mass, 
$r$, could be used to probe the physics of the disk and, perhaps, 
place a limit on the black hole spin. For models in which the
large scale outflow is powered by processes in the accretion disk,
the black hole spin may be studied by combining the beam power of the 
outflow and mass of the black hole.  

There are 19 powerful radio galaxies for which we have estimates of 
both the black hole mass $M$ and the total energy $E_*$. 
All of these very powerful FRII radio sources   
have radio powers at least a 
factor of ten above the classical FRI/FRII transition, 
and are referred to as FRIIb sources. 
The sources and their properties are listed in
Table 1.  The source redshifts range from 0.056 to 1.79, and
the hot spot to hot spot source sizes range from about 100 kpc to 
over 600 kpc (O'Dea et al. 2008). 
The black hole mass for Cygnus A (3C 405) is obtained from 
Tadhunter et al. (2003); those for the next 
four sources are obtained from 
McLure et al. (2004), and those for the remaining 14 sources
are obtained using equation (3) of McLure et al. (2006) where
the values of $M_{sph}$ were provided by McLure (McLure, private
communication 2008).  The total energies are obtained from 
Wan, Daly, \& Guerra (2000), Guerra, Daly, \& Wan (2000), and
O'Dea et al. (2008) and 
converted to the cosmological model adopted here. 
The energy from the two sides of each source 
are combined to obtain the total outflow energy $E_*$, which 
is taken to be twice the weighted mean of 
the outflow energies from each side of the source. 
Deviations of the magnetic field strength of the 
extended radio source 
from the minimum energy value 
only enter $E_*$ through the normalization of this quantity, as discussed
in detail by O'Dea et al. (2008).  The lobes of Cygnus A are
used to normalize the sources, and an offset of 0.25 
of magnetic field strength from the minimum energy value,
appropriate for Cygnus A
(Carilli et al. 1991; Wellman, Daly, \& Wan 1997) has been 
adopted. Note that if there is no offset from minimum energy conditions
in Cygnus A, the empirically determined values of $E_*$ and $r$ 
decreases by about a 
factor of 5 (O'Dea et al. 2008), 
so the values of $j$ decrease by about $\sqrt{5}$ 
from those discussed below. 

The dimensionless ratio of outflow energy to black hole mass, 
$r\equiv E_*/(c^2M)$, is obtained for each source and substituted
into eq. (1) to obtain 
the spin parameter $j$, and the values are listed in Table 1. 
The sources have very similar values of $r$ and $j$, with no
dependence of either parameter on source size or redshift. 
The weighted mean values of $r$ and $j$ for these 19 sources
are  $(1.6 \pm 0.3)\times 10^{-3}$ and $0.12 \pm 0.01$, respectively. 
Interestingly, within the measurement error, 
the values of $r$ and $j$ for each source are consistent with the mean value
of the 19 sources. 
The values of $j$ are shown as a histogram in Figure 1 and 
as a function of black hole mass in Figure 2.  Figure 3 shows the 
total outflow energy as a function of black hole mass.

It would appear that the values of $r$ and $j$ for these FRIIb sources is  
constant, suggesting that the physical state of the system at
the time the outflow is generated is very similar in each of
the sources studied.  This could indicate that the outflow is triggered
when a particular threshold, given by $r \approx 10^{-3}$, is reached. 
A model that includes a threshold of this type is discussed by Meier (1999). 
A threshold for the onset of the outflow is also indicated by a 
comparison of individual source properties with the properties of the 
full population of sources (e.g. Daly \& Guerra 2002; Daly et al. 2008).

\subsection{Results Obtained with Central Dominant Galaxies}

There are now many examples of 
galaxy clusters in which the intracluster medium (ICM) 
has been 
displaced as the result of powerful outflows 
from AGN (e.g. McNamara \& Nulsen 2007; Rafferty et al. 2006).  
The displacement of the ICM provides a measure of the 
total energy that has been pumped into the ICM by the large-scale outflow from 
the AGN associated with the central dominant galaxy (CDG).
If the highly collimated outflow is powered by the
spin energy of the AGN, then the energy of the outflow and the 
mass of the black hole can be combined to solve for the spin $j$ 
of the black hole. Of the 33 galaxy clusters
studied by Rafferty et al. (2006), 29 have estimates  
of both the mass of the black hole associated with the large-scale
outflow and the total energy input to the ICM by the 
AGN.  One source, A1068, does not have an estimate 
of the total energy input to the ICM by the AGN, and three
sources, RBS 797, MACS J1423.8+2404, and 3C401, do not have
estimates of the black hole mass. The remaining 29 sources are 
considered here. 
Almost all of these sources have FRI 
radio source structure or have amorphous radio structure, with 
a few exceptions such as Cygnus A (Birzan et al. 2008).

Equation (1) was used to estimate the spin of each of the 29 
sources for which there is a measure of both the energy $E_*$ and
the black hole mass $M$. 
The energy $E_* = 4 PV$ and black hole mass for each source 
are obtained from 
Rafferty et al. (2006), where the values of $M_{BH,L_K}$ are used
since they are obtained in a manner that is very similar to the black 
hole mass determinations described in section \ref{FRII}.  The values
of $M_{BH,L_K}$ 
listed in Table 3 of Rafferty et al. (2006) had been modified by 
those authors by a factor of
0.35, and this adjustment has been removed. 
This brings the mass estimates for Cygnus A (3C 405) and M84
into good agreement with the independent determinations by  
Tadhunter et al. (2003) and Maciejewski \& Binney (2001), respectively. 
For simplicity, the average value of asymmetric error bars was used. 

The values obtained for the ratio $r$ and 
spin $j$ of each source are listed in Table 2.
The median value of $j$ is 0.026, the average value of $\log{j}$ is 
$-1.53 \pm 0.03$, 
and the values of $j$ range from about 0.001 to 
0.4. A histogram of the values of $\log{(j)}$ is shown in 
Figure 1, and $j$ is shown as a 
function of black hole mass in Figure 2.  Figure 3 shows the 
total outflow energy as a function of black hole mass. 
Clearly, there is a much broader range of $E_*$, $r$, and $j$ for
the CDGs relative to those of powerful classical
double radio galaxies. 
Most of the values of $\log{(j)}$ lie in the range from $10^{-1}$ to 
$10^{-2}$. The values of the energy extracted for 
the 29 CDGs studied include the energy
extracted up to the time of observation, and each source is being observed
at a different stage in its lifetime.  In addition, relativistic plasma
left from a prior outflow event could cause the outflow energy from 
the current outflow event to be overestimated. 
These effects can cause the observed distributions 
of $r$ and $j$ to be quite broad.

\section{DISCUSSION}

The range of values of $r$ and $j$ for the FRIIb radio galaxies
studied is remarkably small, while the range of values of $r$ and 
$j$ for 
outflows associated with CDGs is quite broad.  
There are two key distinctions between these samples that may 
partially 
explain these differences. Firstly, the outflow energy measured for 
a  FRIIb source is an estimate of the total outflow energy 
that will be produced over the entire lifetime of the outflow, whereas
the outflow energy measured for the CDGs is an estimate of the outflow
energy produced by the source up to the time of the observation. Catching 
each CDG outflow at a different phase in the course of its lifetime 
will cause the distribution of outflow energies for CDGs to be broad; this
is not a factor for the FRIIb sources. In addition, 
a prior outflow event
from a CDG could cause the outflow energy of a given event 
to be overestimated. 
Secondly, the FRIIb radio 
galaxies are selected from the flux limited 3CRR catalogue 
(Laing, Riley, \& Longair 1983), 
so only the most powerful sources are detected at each redshift. 
Interestingly, the total outflow energy 
$E_*$ measured for powerful 3CRR 
radio galaxies increases as $(1+z)^{1.8 \pm 0.2}$ 
(O'Dea et al. 2008), while the black hole mass rises as 
$(1+z)^{2.07 \pm 0.76}$ (McLure et al. 2006), so the ratio $r$ 
is independent
of redshift. The fact that lower mass and lower energy black holes are
not observed at high redshift is due to the flux limit of the survey.
The fact that the high mass and high energy sources FRIIb sources 
are not seen at 
low redshift may suggest that the higher mass sources are more active at
high redshift, or that changes in the environment cause the radio power
of the sources to fall below the flux limit of the survey, that is, the 
sources may be lower power FRII sources or FRI sources, though they could
still have large outflow energies (e.g. Hardee et al. 1992;  O'Donoghue, 
Eilek, \& Owen 1993; Bicknell 1995).
A selection bias also seems to be present in the CDG sample; the 6 sources
with outflow energies less than $10^4 c^2 M_{\odot}$ are all at very 
low redshift, and the outflow energy, black hole mass, and black hole
spin all tend to increase with redshift.

Very powerful 3CRR radio galaxies are likely to evolve into the
CDGs (Lilly \& Longair 1984; Best et al. 1998;
McLure et al. 2004). Thus, the two samples studied are representative of the 
early and late phases of evolution of these systems. The systems have
black holes with similar mass.  Clearly, the FRIIb sources have a very
small range of outflow energy per unit black hole mass, $r$.  It would appear
that there is a larger range of values for this parameter for CDGs, 
but, given the factors 
affecting the measurements of outflow energy from black holes of
CDGs, it is possible that these sources have a similarly small
intrinsic range of $r$ and $j$. 

The results obtained here are consistent with 
the idea that the different
radio structures of FRI and FRII sources are due to differences
in their gaseous 
environments (e.g. 
Burns et al. 1994; Bicknell 1995), and suggests that the gaseous
environments of these sources has evolved significantly.  The evolution
of the gaseous environment could be due in part to the large-scale
outflows (e.g. Silk \& Rees 1998; Eilek \& Owen 2002). 

It is remarkable that so few sources have values of 
$j$ close to the theoretical limit of 
unity; that is, almost all sources have values of $r$ much less than 
the theoretical limit of 0.29. 
All of the FRIIb sources have $r \sim 10^{-3}$.  
This suggests that each system is in a similar physical state at the 
time the outflow is generated, and may indicate that the outflow is
triggered when a particular threshold is reached.  
If the sources have intrinsic values 
of $j$ that start out close to one, 
each outflow event is tapping a very small fraction 
of the available spin energy, and each source may  
have multiple outflow events. 

The spin values obtained here could be compared with independent measures
of the black hole spin.  This may provide a diagnostic of 
whether it is the black hole
spin or the accretion disk 
that is powering large-scale outflows from these sources. 
In addition,  
the method of using outflow energies and black hole masses to estimate
black hole spins presented here 
may be applicable to other systems. 

\acknowledgements
It is a pleasure to thank George Djorgovski,  Brian McNamara, 
David Meier, and 
the referee of this paper  
for very helpful comments and suggestions on this work. It is 
also a pleasure to thank Ross McLure for providing stellar masses for
the fourteen highest redshift sources listed in Table 1. 
This work was supported in part by U. S. National Science
Foundation grants AST-0507465.

\begin{deluxetable}{lclcll}
\tablewidth{0pt}
\tablecaption{FRIIb Black Hole Properties}
\tablehead{
\colhead{Source} &\colhead{z} & \colhead{$E_*/c^2$ } &  \colhead{$M$} 
&\colhead{$r\equiv E_*/(c^2M)$} &\colhead{$j$}\\
&&\colhead{$(10^6 M_{\odot})$}&\colhead{$(10^8 M_{\odot}$)}
&\colhead{$(10^{-3})$}}
\startdata  
3C 405	&$	0.056	$&$	3.2	\pm	0.6	$&$	25	\pm	7	$&$	1.3	\pm	0.4	$&$	0.10	\pm	0.02	$\\
3C 244.1	&$	0.43	$&$	1.8	\pm	0.3	$&$	9.5	\pm	6.6	$&$	1.9	\pm	1.4	$&$	0.12	\pm	0.04	$\\
3C 172	&$	0.519	$&$	2.8	\pm	0.6	$&$	7.8	\pm	5.4	$&$	3.6	\pm	2.6	$&$	0.17	\pm	0.06	$\\
3C 330	&$	0.549	$&$	4.3	\pm	0.9	$&$	13	\pm	9	$&$	3.3	\pm	2.4	$&$	0.16	\pm	0.06	$\\
3C 427.1	&$	0.572	$&$	2.6	\pm	0.6	$&$	15	\pm	10	$&$	1.8	\pm	1.3	$&$	0.12	\pm	0.04	$\\
3C 337	&$	0.63	$&$	2.3	\pm	0.5	$&$	9.1	\pm	6.2	$&$	2.5	\pm	1.8	$&$	0.14	\pm	0.05	$\\
3C34	&$	0.69	$&$	3.8	\pm	0.7	$&$	16	\pm	11	$&$	2.4	\pm	1.7	$&$	0.14	\pm	0.05	$\\
3C441	&$	0.707	$&$	4.1	\pm	0.7	$&$	18	\pm	12	$&$	2.3	\pm	1.7	$&$	0.14	\pm	0.05	$\\
3C 55	&$	0.735	$&$	6.2	\pm	1.3	$&$	14	\pm	10	$&$	4.3	\pm	3.2	$&$	0.18	\pm	0.07	$\\
3C 247	&$	0.75	$&$	3.0	\pm	0.6	$&$	26	\pm	18	$&$	1.2	\pm	0.9	$&$	0.10	\pm	0.04	$\\
3C 289	&$	0.9674	$&$	4.3	\pm	0.9	$&$	27	\pm	21	$&$	1.6	\pm	1.3	$&$	0.11	\pm	0.05	$\\
3C 280	&$	0.996	$&$	3.6	\pm	0.7	$&$	27	\pm	21	$&$	1.3	\pm	1.1	$&$	0.10	\pm	0.04	$\\
3C 356	&$	1.079	$&$	7.6	\pm	1.8	$&$	28	\pm	22	$&$	2.7	\pm	2.3	$&$	0.15	\pm	0.06	$\\
3C 267	&$	1.142	$&$	6.5	\pm	1.3	$&$	24	\pm	20	$&$	2.6	\pm	2.2	$&$	0.15	\pm	0.06	$\\
3C 324	&$	1.206	$&$	5.8	\pm	1.4	$&$	37	\pm	30	$&$	1.6	\pm	1.4	$&$	0.11	\pm	0.05	$\\
3C 437	&$	1.48	$&$	13	\pm	3	$&$	24	\pm	22	$&$	5.5	\pm	5.1	$&$	0.21	\pm	0.10	$\\
3C 68.2	&$	1.575	$&$	6.8	\pm	1.5	$&$	35	\pm	32	$&$	2.0	\pm	1.9	$&$	0.13	\pm	0.06	$\\
3C 322	&$	1.681	$&$	10.6	\pm	2.2	$&$	32	\pm	30	$&$	3.3	\pm	3.2	$&$	0.16	\pm	0.08	$\\
3C 239	&$	1.79	$&$	10.2	\pm	3	$&$	37	\pm	36	$&$	2.8	\pm	2.8	$&$	0.15	\pm	0.07	$\\
\enddata
\label{Table}
\end{deluxetable}

\begin{deluxetable}{lccccc}
\tablewidth{0pt}
\tablecaption{CDG Black Hole Properties}
\tablehead{
\colhead{Source} &\colhead{z} & \colhead{$E_*/c^2$ } &  \colhead{$M$} 
&\colhead{$r\equiv E_*/(c^2M)$} &\colhead{$j$}\\
&&\colhead{$(10^6 M_{\odot})$}&\colhead{$(10^8 M_{\odot}$)}
&\colhead{$(10^{-3})$}}
\startdata
        MS 0735.6+7421	&	0.216	&$	36	\pm	26	$&$	20	\pm	11	$&$	18	\pm	16	$&$	0.37	\pm	0.16	$\\
        Zw 2701     	&	0.214	&$	7.8	\pm	8.1	$&$	17	\pm	9	$&$	4.5	\pm	5.2	$&$	0.19	\pm	0.11	$\\
        Hydra A     	&	0.055	&$	1.4	\pm	0.7	$&$	11	\pm	4	$&$	1.2	\pm	0.7	$&$	0.10	\pm	0.03	$\\
        Zw 3146     	&	0.291	&$	8.4	\pm	6.3	$&$	74	\pm	53	$&$	1.1	\pm	1.2	$&$	0.10	\pm	0.05	$\\
        MKW 3S      	&	0.045	&$	0.84	\pm	0.48	$&$	8.6	\pm	2.9	$&$	0.99	\pm	0.65	$&$	0.089	\pm	0.029	$\\
        Cygnus A    	&	0.056	&$	1.9	\pm	0.9	$&$	29	\pm	14	$&$	0.65	\pm	0.46	$&$	0.072	\pm	0.025	$\\
        PKS 0745-191	&	0.103	&$	1.5	\pm	0.7	$&$	31	\pm	16	$&$	0.49	\pm	0.34	$&$	0.062	\pm	0.022	$\\
        Hercules A  	&	0.154	&$	0.69	\pm	0.54	$&$	20	\pm	11	$&$	0.34	\pm	0.34	$&$	0.052	\pm	0.026	$\\
        Sersic 159/03	&	0.058	&$	0.56	\pm	0.38	$&$	17	\pm	9	$&$	0.32	\pm	0.27	$&$	0.051	\pm	0.021	$\\
        A133        	&	0.060	&$	0.53	\pm	0.13	$&$	20	\pm	10	$&$	0.27	\pm	0.15	$&$	0.046	\pm	0.013	$\\
        Perseus     	&	0.018	&$	0.42	\pm	0.28	$&$	17	\pm	7	$&$	0.25	\pm	0.19	$&$	0.044	\pm	0.017	$\\
        A1835       	&	0.253	&$	1.0	\pm	0.7	$&$	54	\pm	36	$&$	0.19	\pm	0.19	$&$	0.039	\pm	0.019	$\\
        4C 55.16    	&	0.242	&$	0.27	\pm	0.18	$&$	14	\pm	7	$&$	0.19	\pm	0.16	$&$	0.039	\pm	0.016	$\\
        A2597       	&	0.085	&$	0.080	\pm	0.068	$&$	8.6	\pm	2.9	$&$	0.093	\pm	0.085	$&$	0.027	\pm	0.012	$\\
        A2199       	&	0.030	&$	0.17	\pm	0.09	$&$	20	\pm	9	$&$	0.083	\pm	0.057	$&$	0.026	\pm	0.009	$\\
        3C 388      	&	0.092	&$	0.12	\pm	0.11	$&$	17	\pm	7	$&$	0.067	\pm	0.068	$&$	0.023	\pm	0.012	$\\
        A1795       	&	0.063	&$	0.10	\pm	0.09	$&$	23	\pm	11	$&$	0.046	\pm	0.046	$&$	0.019	\pm	0.010	$\\
        A4059       	&	0.048	&$	0.067	\pm	0.038	$&$	29	\pm	14	$&$	0.023	\pm	0.018	$&$	0.014	\pm	0.005	$\\
        A2052       	&	0.035	&$	0.038	\pm	0.033	$&$	17	\pm	7	$&$	0.022	\pm	0.022	$&$	0.013	\pm	0.006	$\\
        A2029       	&	0.077	&$	0.11	\pm	0.03	$&$	60	\pm	36	$&$	0.018	\pm	0.012	$&$	0.012	\pm	0.004	$\\
        2A 0335+096 	&	0.035	&$	0.024	\pm	0.014	$&$	14	\pm	7	$&$	0.017	\pm	0.013	$&$	0.012	\pm	0.005	$\\
        A478        	&	0.081	&$	0.033	\pm	0.017	$&$	26	\pm	14	$&$	0.013	\pm	0.010	$&$	0.010	\pm	0.004	$\\
        A85         	&	0.055	&$	0.027	\pm	0.018	$&$	29	\pm	14	$&$	0.0093	\pm	0.0078	$&$	0.0086	\pm	0.0036	$\\
        PKS 1404-267	&	0.022	&$	0.0027	\pm	0.0022	$&$	5.7	\pm	2.9	$&$	0.0047	\pm	0.0045	$&$	0.0061	\pm	0.0030	$\\
        A262        	&	0.016	&$	0.0029	\pm	0.0014	$&$	8.6	\pm	2.9	$&$	0.0034	\pm	0.0020	$&$	0.0052	\pm	0.0016	$\\
        HCG 62      	&	0.014	&$	0.0010	\pm	0.0011	$&$	5.7	\pm	2.9	$&$	0.0018	\pm	0.0022	$&$	0.0038	\pm	0.0023	$\\
        Centaurus   	&	0.011	&$	0.0013	\pm	0.0007	$&$	8.6	\pm	2.9	$&$	0.0016	\pm	0.0010	$&$	0.0035	\pm	0.0011	$\\
        M87         	&	0.0042	&$	0.00044	\pm	0.00019	$&$	8.6	\pm	2.9	$&$	0.00052	\pm	0.00028	$&$	0.0020	\pm	0.0006	$\\
        M84         	&	0.0035	&$	0.000067	\pm	0.000078	$&$	3.4	\pm	0.9	$&$	0.00019	\pm	0.00023	$&$	0.0012	\pm	0.0007	$\\
\enddata
\label{Table}
\end{deluxetable}

\begin{figure}
    \centering
    \includegraphics[width=\textwidth]{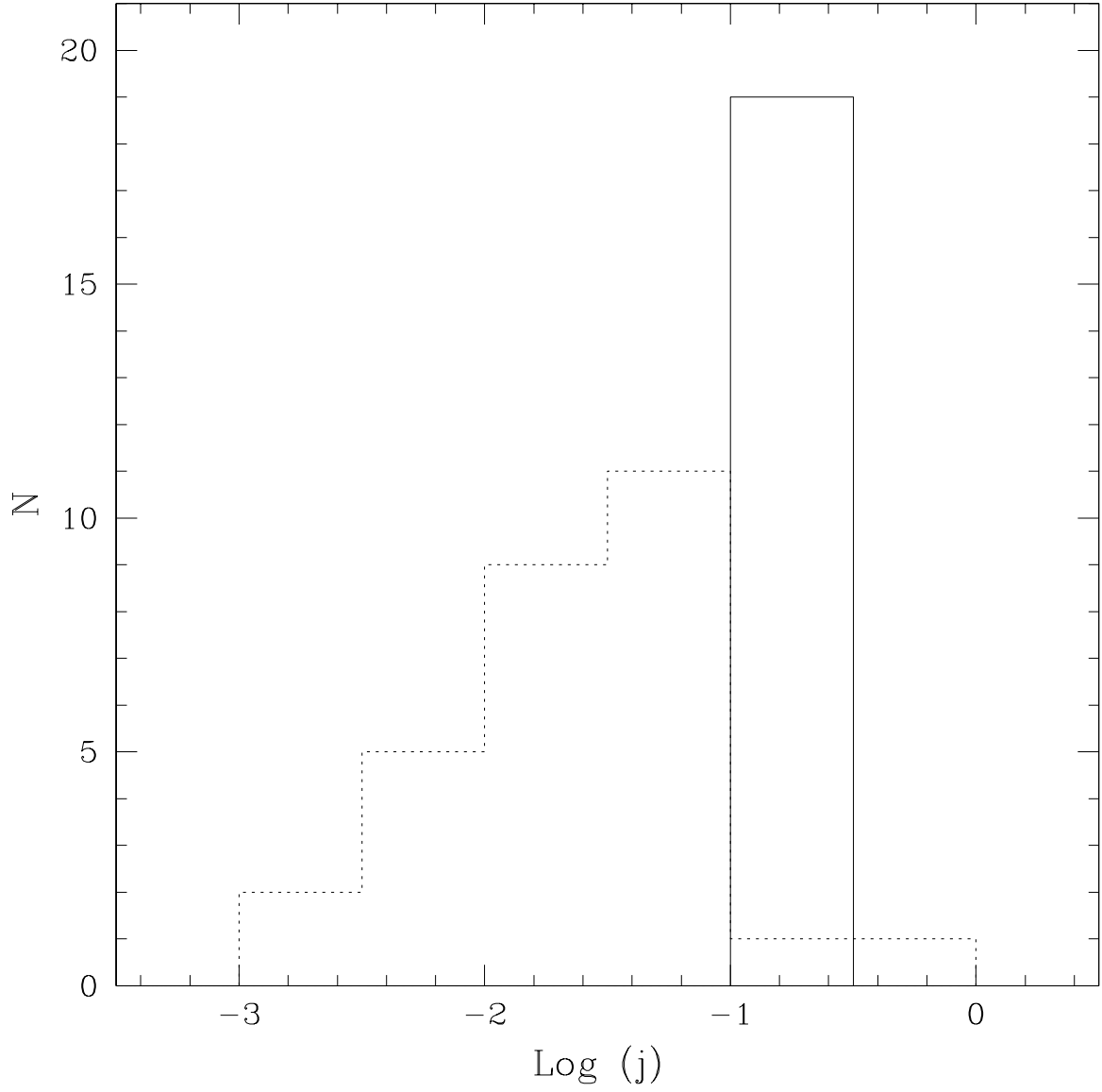}
       \caption{Histogram of spin values.  The dotted
line indicated the histogram for the sample of 29 central dominant
galaxies, while the solid line indicates that for the sample of
19 powerful classical double radio galaxies.  }
          \label{fighist}
    \end{figure}
\begin{figure}
    \centering
    \includegraphics[width=\textwidth]{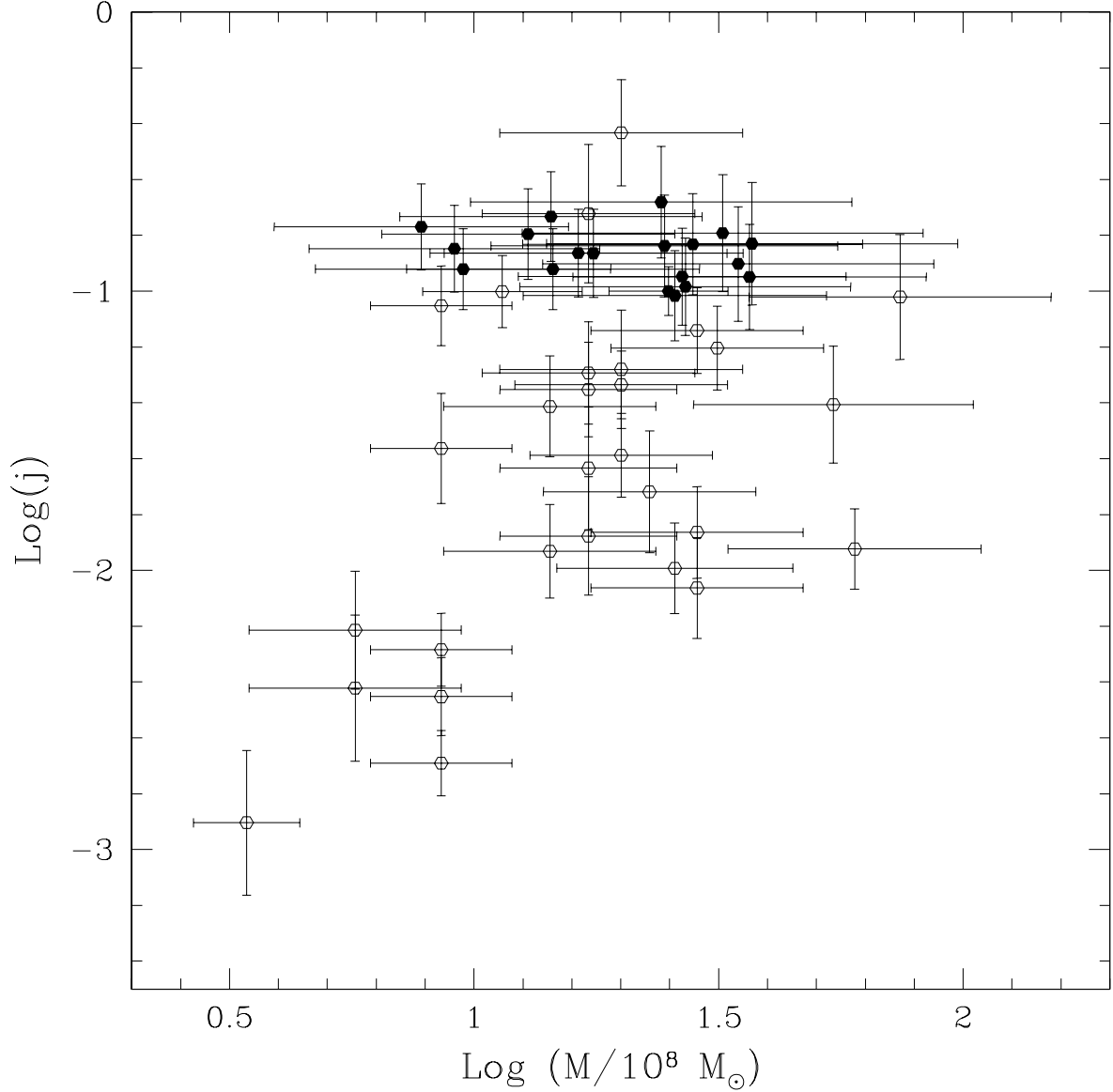}
       \caption{Distribution of black hole spin as a 
function of black hole mass.  
The 19 sources associated 
with very powerful classical double 
radio galaxies are indicated by solid 
circles, and the 29 sources associated with 
CDGs are indicated by open circles. 
One source, Cygnus A (3C 405) is included
in both samples.  }
          \label{figjofM}
    \end{figure}

\begin{figure}
    \centering
    \includegraphics[width=\textwidth]{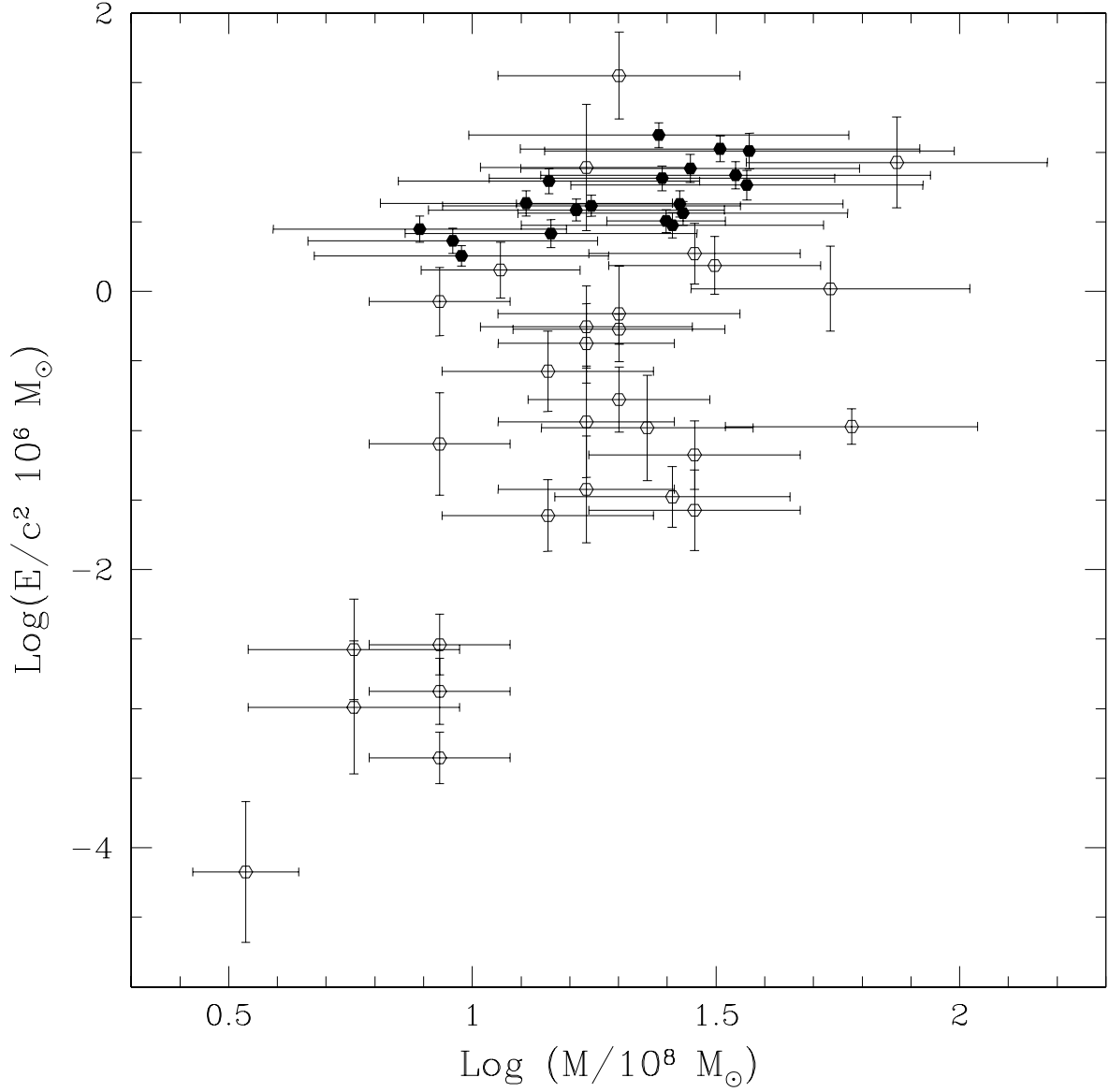}
       \caption{Distribution of outflow energy as a function of
black hole mass; 
the symbols are as in Fig. \ref{figjofM}. }
          \label{figEofM}
    \end{figure}

\end{document}